\documentclass{raa}
\usepackage{graphicx,times}
\usepackage{natbib}
\usepackage{amssymb,amsmath}
\bibpunct{(}{)}{;}{a}{}{,}

\usepackage[pagebackref=true]{hyperref}
\hypersetup{colorlinks = true, linkcolor = green, anchorcolor = red, citecolor = blue,
 filecolor = red, pagecolor = red, urlcolor = red}

\begin{document}

\title{Polarised radio emission associated with HESS J1912+101}
\subtitle {}
\author{W.~Reich\inst{1}, X. H. Sun\inst{2}}

\offprints{wreich@mpifr-bonn.mpg.de}

\institute{
  Max-Planck-Institut f\"{u}r Radioastronomie, Auf dem H\"{u}gel 69,
  53121 Bonn, Germany {\it wreich@mpifr-bonn.mpg.de}\\
\and Department of Astronomy, Yunnan University, and Key Laboratory of Astroparticle
Physics of Yunnan Province, Kunming, 650091, People's republic of China\\
\vs \no
   {\small Received 20XX Month Day; accepted 20XX Month Day}
}


\abstract
{The shell-type TeV source HESS J1912+101 was tentatively identified as an 
old supernova remnant, but is missing counterparts at radio and other frequencies. 
We analysed the Sino-German Urumqi $\lambda$6\ cm survey and the Effelsberg 
$\lambda$11\ cm and $\lambda$21\ cm surveys to identify radio emission from 
HESS J1912+101 to clarify the question of a supernova origin.  
We find a partial shell of excessive polarisation at $\lambda$6\ cm at the periphery
of HESS J1912+101. At $\lambda$11\ cm, its polarised emission is faint and suffers 
from depolarisation, while at $\lambda$21\ cm, no related polarisation is seen.
We could not separate the shell's total intensity signal from the confusing 
intense diffuse emission from the inner Galactic plane. However, a high-percentage 
polarisation of the shell's synchrotron emission is indicated. 
Our results support earlier suggestions that HESS J1912+101 is an old supernova 
remnant. The synchrotron emission is highly polarised, which is typical for 
evolved supernova remnants of low surface-brightness.  
}
\authorrunning{W. Reich and X. H. Sun}
\titlerunning{Polarisation towards HESS J1912+101}

\maketitle
\section{Introduction}

SNR shock-fronts are major cosmic-ray acceleration sites in the Galaxy,
where X-ray and TeV-emission indicates acceleration up to very high 
energies. The catalogue of Galactic supernova remnants (SNRs) by \citet{Green17} 
currently lists 295 objects. From the known SNRs, only a very small fraction is 
not visible in the radio range. Estimates of the total number of existing SNRs 
in the Galaxy predict about three times more objects than known today 
\citep{Li91,Tammann94}, which 
results from observational selection effects. 
Compact, young SNRs at large distance and evolved, faint, extended objects are hidden in 
the intense emission along the Galactic plane, where most SNRs are located. In recent years,
some TeV emitting SNR candidates were discovered by HESS with no or faint counterparts
in the X-ray or the radio range \citep{HESS2018} indicating a preference for
high-energy cosmic rays, which is unexpected. With increasing sensitivity, TeV telescopes
will likely find more such objects. Their emission or upper limits at other wavelenths 
is needed to understand these sources.         

HESS J1912+101 is an extended TeV object first reported by \citet{Aharonian08},
and recently, with more sensitive data, was shown to resemble a thick 
shell-type object \citep{Puehlhofer15, Gottschall17, HESS2018}.
HESS J1912+101 is located in the Galactic plane with Galactic coordinates 
of $l,b = 44\fdg46, -0\fdg13$ at its centre, which means a Galactic designation 
G44.46-0.13. The shape is slightly elliptical with an extent of  
approximately $50\arcmin \times 58\arcmin$. HESS J1912+101 was
tentatively interpretated as an old SNR \citep{Puehlhofer15, Gottschall17, HESS2018}, 
although this interpretation remains inconclusive because of missing signatures at 
other wavelengths.

Recently, \citet{Su17} supported the SNR interpretation for HESS J1912+101 based on 
the identification of associated shocked molecular gas and high-velocity HI. \citet{Su17} 
derived a distance of 4.1~kpc to HESS J1912+101. These authors have also analysed 
the spectrum of HESS J1912+101 and used the model of \citet{Yamazaki06} for the emission 
of TeV ${\gamma}$-rays from evolved SNRs to estimate its radio emission. This model 
predicts a high ratio of TeV to radio flux density for $10^5$ 
year old SNRs. Thus, \citet{Su17} estimated a 1.4~GHz flux density of HESS J1912+101 
in the range of 0.1~Jy to 0.7~Jy. For the putative radio shell of HESS 
J1912+101  this means a surface brightness at 1~GHz below $3.6\times10^{-23}$~Wm$^{-2}$Hz$^{-1}$sr$^{-1}$,
which is among the lowest SNR surface brightness values currently known. Such
faint SNRs were all identified in low-emission areas, either well outside of the 
Galactic plane, e.g. G181.1+9.5 \citep{Kothes17}, G156.2+5.7 \citep{Reich92, Xu07}, 
G65.2+5.7 \citep{Reich79, Xiao09} or in the Galactic anticentre region,
e.g. G152.4-2.1 and G190.9-2.2 \citep{Foster13} or G178.2-4.2 \citep{Gao11y}. 
As already mentioned, these faint SNRs are extremely difficult to find as discrete 
shell-type sources within the high level of confusing diffuse emission in the inner 
Galactic plane. Indeed, associated radio synchrotron emission has not been identified 
for HESS J1912+101 from available VLA continuum surveys \citep{Puehlhofer15, HESS2018}.
A possible association with the SNR candidate from the Clark Lake 30.9~MHz survey, 
G44.6+0.1, with a quoted flux density of 19~Jy \citep{Kassim881, Kassim882, Gorham90}
was discussed by \citet{HESS2018}. This Clark Lake source has no counterpart matching 
in strength and extent at any other radio frequency, what raises doubts on its 
reality. 

The known faint low-surface brightness SNRs 
often show a very high percentage polarisation of the order of {50\%} or higher. 
Polarisation suffers less from confusion than total intensities, but depolarisation lowers 
the polarised signal from SNRs if they are too far away, the observing frequency is not 
high enough, or the angular resolution is too coarse. It is therefore of interest to check 
the area of HESS J1912+101 for possibly related polarised radio emission as an alternative
to settle its SNR identification by the synchrotron emission process.   

\section{Data extraction}

\subsection{The Urumqi $\lambda$6\ cm and the WMAP K-band survey}

We make use of the Sino-German $\lambda$6\ cm (4800~MHz) polarisation survey of the Galactic plane,
which is the ground-based survey at the highest frequency currently available \citep{Han15}.
The survey was carried out with the Urumqi 25-m radio telescope of Xinjiang 
Astronomical Observatories, Chinese Academy of Sciences, between 2004 and 2009 and covers 
the northern Galactic plane for longitudes from $10\degr$ to $230\degr$ and latitudes 
of $\pm5\degr$. 
The angular resolution of the survey is 9$\farcm$5. The survey's concept, its observation 
and the reduction and calibration procedures were discussed in some detail by \citet{Sun07}.
The survey was published in three sections by \citet {Gao10, Xiao11, Sun11a}. 
The Urumqi $\lambda$6\ cm survey maps are available for download
\footnote{http://zmtt.bao.ac.cn/6cm/}. $\lambda$6\ cm flux densities of 3823 compact 
sources were published by \citet {Reich14}.   

The $\lambda$6\ cm data of the area around HESS J1912+101 were extracted 
from the Urumqi survey section published by \citet{Sun11a}, where we re-calculated 
the absolute polarisation level for the Urumqi data by using the meanwhile available 
9-year release of the WMAP K-band $\lambda$1.3\ cm survey \citep{Bennett13} following 
the procedure described by \citet{Sun07} and using the spectral indices for extrapolation 
by \citet{Sun11a}. The sensitivity for this survey section is 1~mK T$_{b}$ for total 
intensities and 0.5~mK T$_{b}$ for polarised intensities.
All Urumqi $\lambda$6\ cm survey maps or selected fields in Galactic or Equatorial
coordinates with or without 9-year WMAP polarisation corrections are available from the 
MPIfR survey sampler
\footnote{http://www.mpifr-bonn.mpg.de/survey.html}.

We also use the low-resolution WMAP K-band (22.8~GHz) data \citep{Bennett13} to directly 
show the polarisation properties in the HESS J1912+101 area almost free from 
Faraday rotation effects. 

\subsection{The Effelsberg $\lambda$11\ cm survey}

We have used data from the Effelsberg $\lambda$11\ cm (2695~MHz) inner Galactic plane 
survey \citep{Reich84, Reich9011} and the polarisation maps \citep{Junkes87, Duncan99}
for a search of related radio emission in the field of HESS J1912+101. The Effelsberg 
$\lambda$11\ cm survey has an angular resolution of 4$\farcm$3, a rms-sensitvity of
20~mK T$_{b}$ for total intensities and 11~mK T$_{b}$ for polarised intensities
and is available for download like the Urumqi $\lambda$6\ cm survey and many other 
continuum and polarisation surveys from the MPIfR survey sampler$^2$. 

\subsection{The Effelsberg $\lambda$21\ cm survey}

We also inspected data from an unpublished section of the Effelsberg Medium Latutude 
Survey (EMLS) at $\lambda$21\ cm (1.4~GHz). The EMLS covers the northern 
Galactic plane within {$\pm 20\degr$} Galactic latitude \citep{Uyaniker98, Uyaniker99, Reich04}
and has an angular resolution of 9$\farcm$4. The EMLS maps are on an absolute zero-level,
where $\lambda$21\ cm data from the Stockert survey \citep{Reich82, Reich86} and
the DRAO survey \citep{Wolleben06} were added to total intensities and the polarisation data, 
respectively. The rms-sensitivity for total intensities is 15~mK T$_{b}$ 
and limited by confusion, and about 8~mK T$_{b}$ for Stokes $U$ and $Q$.

\section{Results}

\subsection{Polarised intensities}

\subsubsection{Overview at $\lambda$6\ cm and $\lambda$1.3\ cm}

\begin{figure} 
\centering
\includegraphics[angle=0, width=0.9\textwidth]{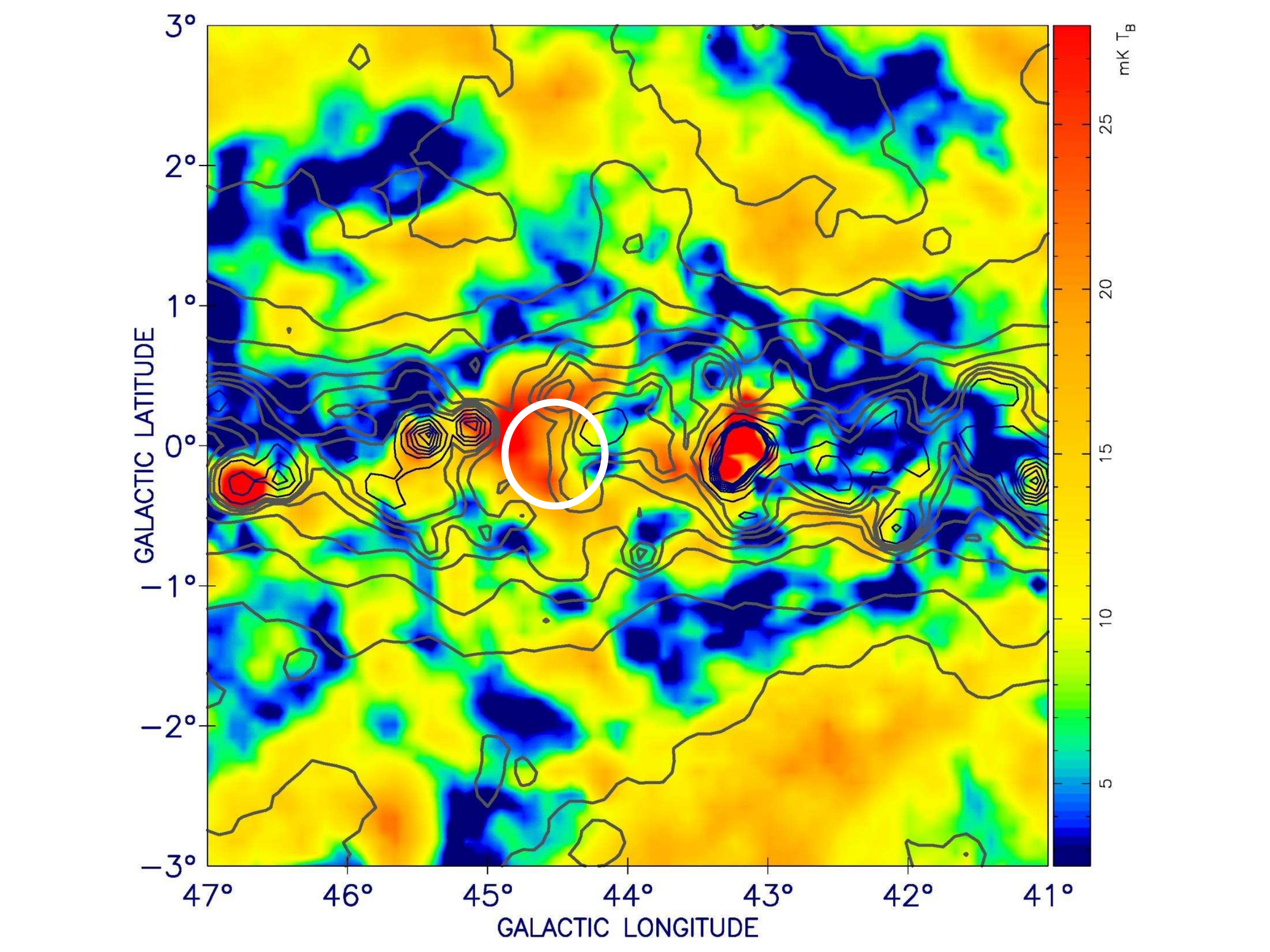}
\caption{Section of the Urumqi $\lambda$6\ cm survey showing polarised intensities 
   including WMAP based baselevel corrections. Selected $\lambda$6\ cm total-intensity 
   contours are overlaid running in steps of 50~mK T$_{b}$ starting at 
   25~mK T$_{b}$.  The bright SNRs W49B ($l,b = 43\fdg2,-0\fdg2$),
   and HC30 ($l,b = 46\fdg8,-0\fdg3$) stand out in polarisation.  
   The approximate boundary of the TeV emission ($\ge 3\sigma$) from 
   HESS J1912+101 is marked by a white circle.}
\label{6cmlarge}
\end{figure}

Figure~\ref{6cmlarge} gives an overview of the $\lambda$6\ cm polarised intensities 
of a 6$\degr \times 6\degr$ section of the Galactic plane with HESS J1912+101 located
in its centre. The Urumqi survey map includes extrapolated large-scale data from the 
WMAP 9-year K-band ($\lambda$1.3\ cm) all-sky survey. 
Polarised and total intensities show the two known, very bright SNRs in the field, 
W49B and HC30, very clearly. Spurious instrumental polarisation close to strong 
sources is also visible. In addition, we note an outstanding polarised 
feature running along the eastern and north-eastern boundary of HESS J1912+101 as marked in 
Fig.~\ref{6cmlarge}. In this area, the arc is the strongest polarisation feature, 
next in strength to the polarised emission from the two known SNRs. The polarised arc is 
not related to any of the thermal HII-regions nearby, which are all very distant
\citep{Anderson12}.
We argue in the following that this polarised arc or partial shell is related to 
HESS J1912+101 and indicates the existence of related synchrotron emission.

\begin{figure} 
\centering
\includegraphics[angle=0, width=0.9\textwidth]{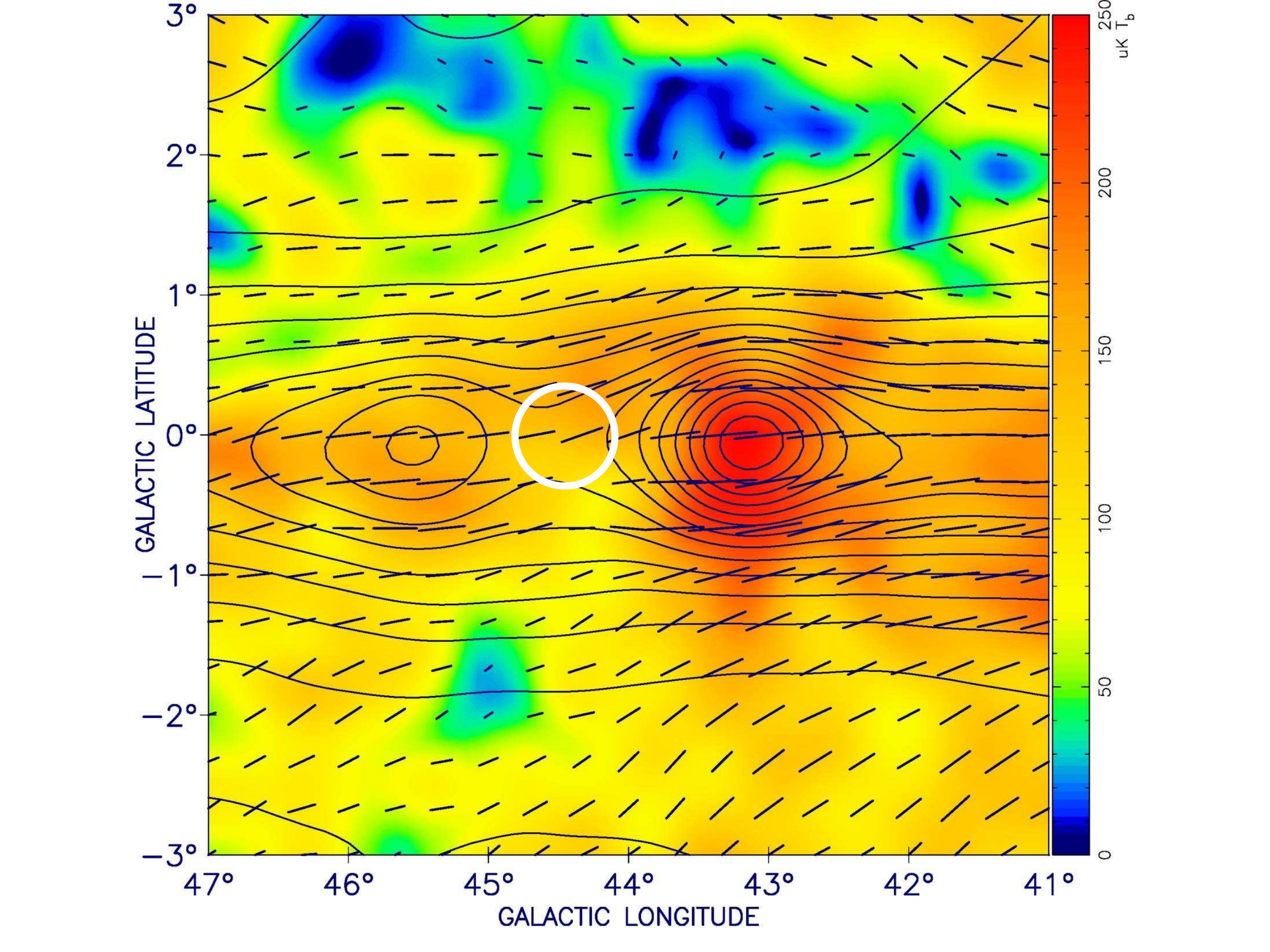}
\caption{The same area as shown in Fig.~1 from the WMAP $\lambda$1.3\ cm survey 
   (9-yr release) showing polarised intensities and selected total-intensity 
   contours. The contours run in steps of 20~$\mu$K T$_{b}$ starting at 
   20~$\mu$K T$_{b}$. At $60\arcmin$ angular resolution, source complexes seen
   in Fig.~1 remain unresolved. Polarised emission from the W49 area is  
   visible. The polarisation vectors are in B-field direction.
   The approximate boundary of HESS J1912+101 is marked by a white circle.}
\label{K9large}
\end{figure}

Figure~\ref{K9large} shows the corresponding WMAP $\lambda$1.3\ cm (K-band) polarisation 
data for the area of Fig.~\ref{6cmlarge}.
The WMAP K-band data were convolved from $52\farcm8$ to $60\arcmin$ to enhance 
the signal-to-noise ratio. Polarised emission from the W49 area including W49B 
is visible, while polarised emission from HC30 is too faint to contrast from
the diffuse emission. 
The diffuse polarised emission increases towards the Galactic plane and reaches 
a typical level of 150$\pm20~\mu$K T$_{b}$, which is also seen towards HESS J1912+101.

For a spectral index of $\beta$ = -2.7 (T$_{b} \sim \nu^{\beta}$), which \citet{Sun11a} 
found to be typical for polarised emission from this area, the expected polarised 
intensity at $\lambda$6\ cm then is about 10$\pm1.5$~mK T$_{b}$,
which is indeed observed in some areas along the Galactic plane. However, a 
level near 10~mK T$_{b}$ is only expected if the depolarisation properties 
at $\lambda$6\ cm are the same as at $\lambda$1.3\ cm, which is unlikely  
the case everywhere and thus lower $\lambda$6\ cm polarised intensities are seen
as well in Fig.~\ref{6cmlarge}. 
The $\lambda$6\ cm polarised emission from the arc at the periphery of HESS J1912+101 
including the diffuse polarised Galactic plane emission is about three times higher than 
expected from the K-band data. 

We consider the possibility that a Faraday screen (FS) along the 
line-of-sight may have caused the excessive polarised emission from the polarised
arc, which would make it an enhancement originating from a superposition of
diffuse Galactic emission components along the line-of-sight and thus the
coincidence with HESS J1912+101 is by chance. 

Observed FSs in general cause a reduction of the polarised intensity 
compared to their surroundings. Foreground and background polarisation angles 
are similarly orientated parallel to the Galactic plane as expected from the 
large-scale Galactic magnetic field direction. Numerous FSs of this type were observed 
and analysed in the Urumqi survey publicatons \citep {Sun07, Gao10, Xiao11, Sun11a}. 

For the case that the foreground and background polarisation angles of a FS differ 
significantly and the FS rotates the background polarisation angle to align 
with the foreground angle, the polarised emission will exceed its surroundings. 
This scenario is possible, but valid for a narrow wavelength range only. 
For the polarised arc, this FS scenario requires that the magnetic field direction 
of the Galactic emission beyond the FS is inclined to the Galactic plane by
70$\degr$ at least, when being very local, and up to 90$\degr$, if more distant.  
However, the K-band polarisation angles, which were almost not affected 
by the rotation measure (RM) of the putative FS, because of the $\lambda^2$ dependence
of RM, do not show large deviations from the orientation along 
the Galactic plane (Fig.~\ref{K9large}), so that the FS scenario can be ruled out.

We conclude that the maximum polarized Galactic $\lambda$6\ cm signal is that extrapolated   
from the K-band polarised emission. At K-band, Faraday rotation by a FS is not important, 
because too strong magnetic fields and thermal densities are required for sufficiently high 
rotation measures to have an effect on the K-band polarisation angles. Thus, the high 
$\lambda$6\ cm polarised emission from the arc can not result from a FS, 
which in turn means that its origin is excessive synchrotron emission.

\begin{figure*}[t]
\includegraphics[angle=0, width=0.5\textwidth]{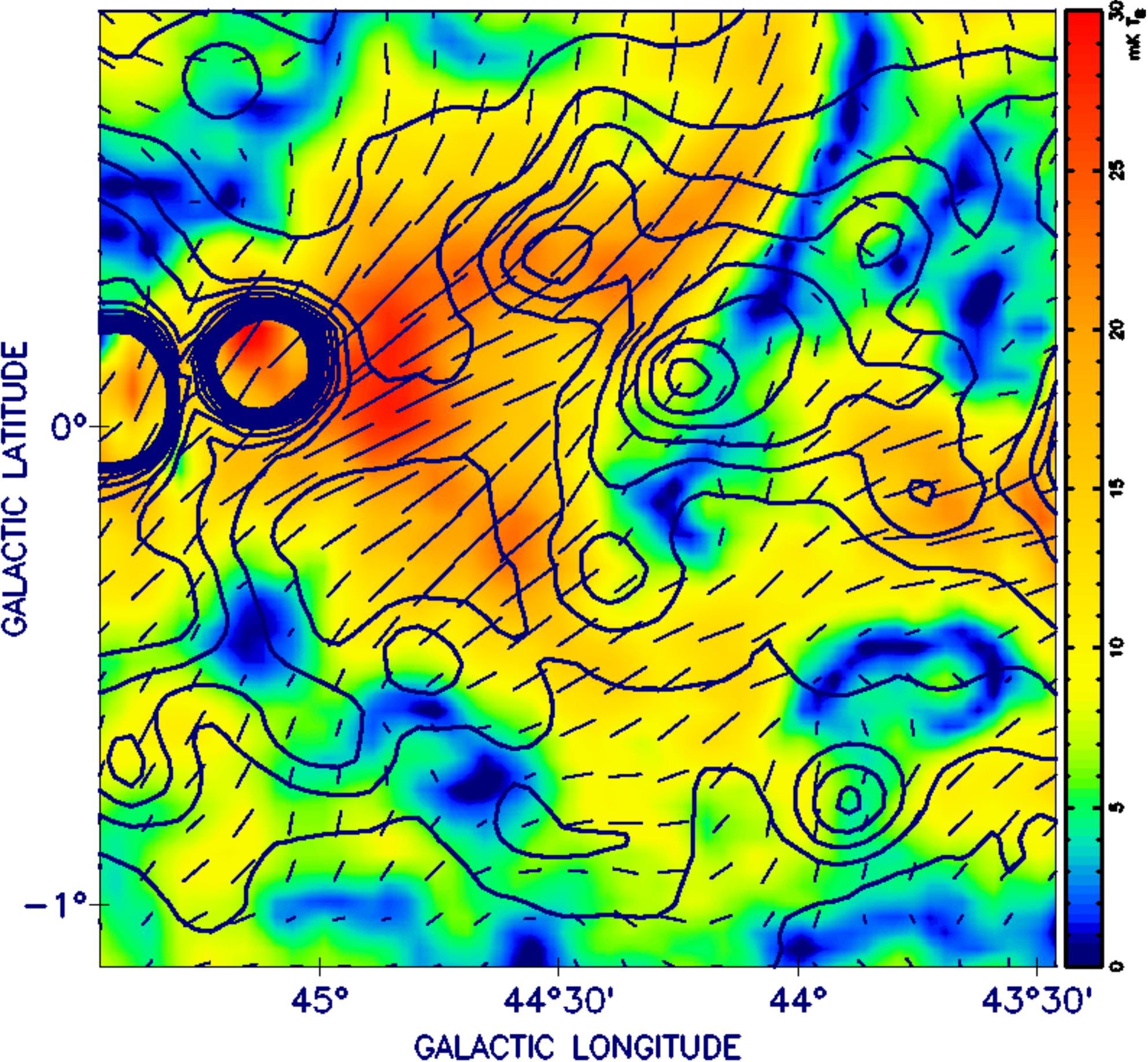}
\includegraphics[angle=0, width=0.5\textwidth]{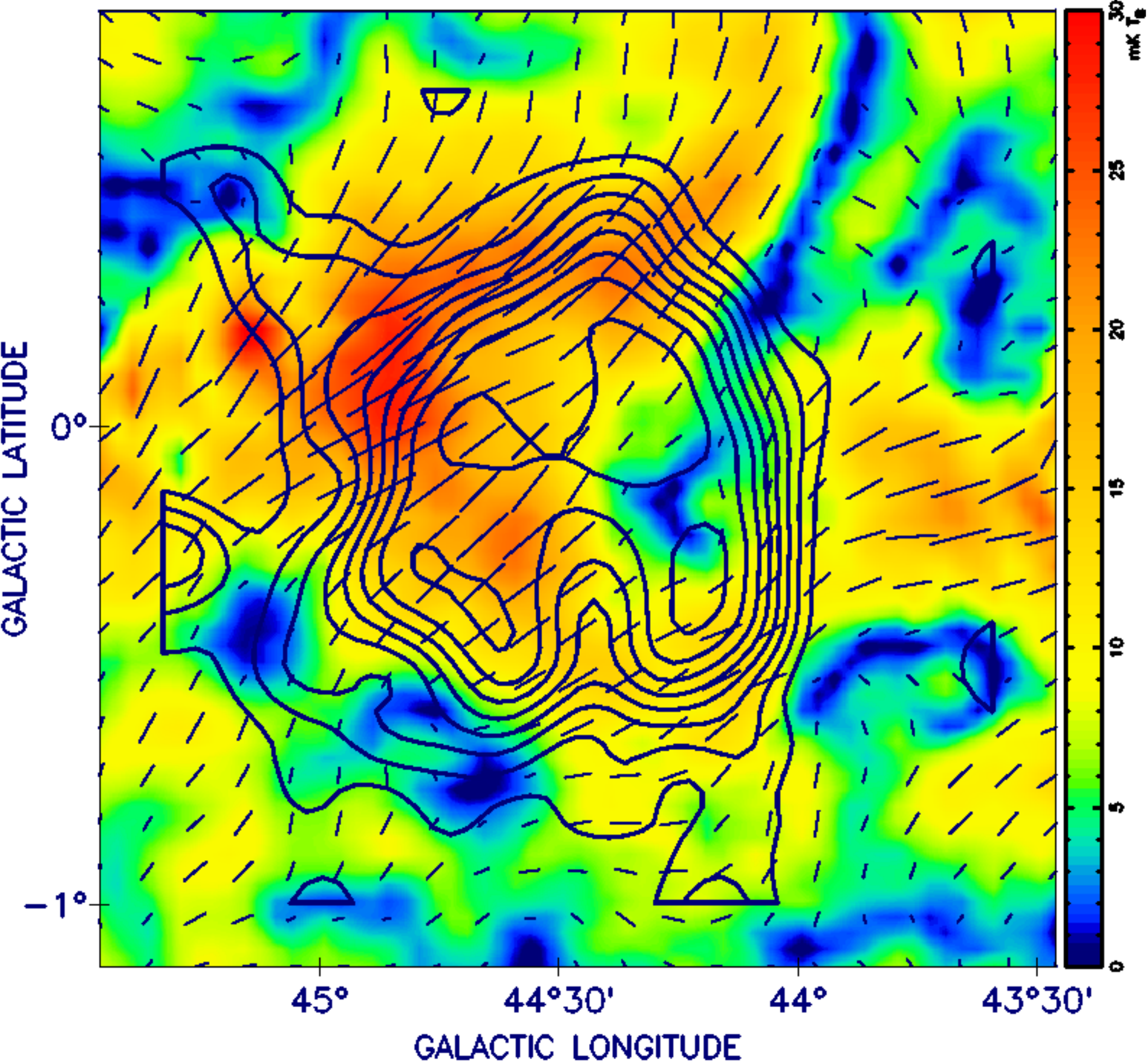}
\caption{{\it Left panel}: Urumqi $\lambda$6\ cm polarised intensity overlaid with 
   polarisation 
   vectors (B-field direction), where the vector lengths is proportional to
   polarised intensity. Overlaid total-intensity contours run in steps of 
   50~mK T$_{b}$ and start at 100~mK T$_{b}$. 
   {\it Right panel}: Urumqi $\lambda$6\ cm polarised intensity and vectors 
   as in the left panel, but overlaid with selected TeV-emission contours from HESS J1912+101.}
\label{6cm}
\end{figure*}

\begin{figure*}
\includegraphics[angle=0, width=0.5\textwidth]{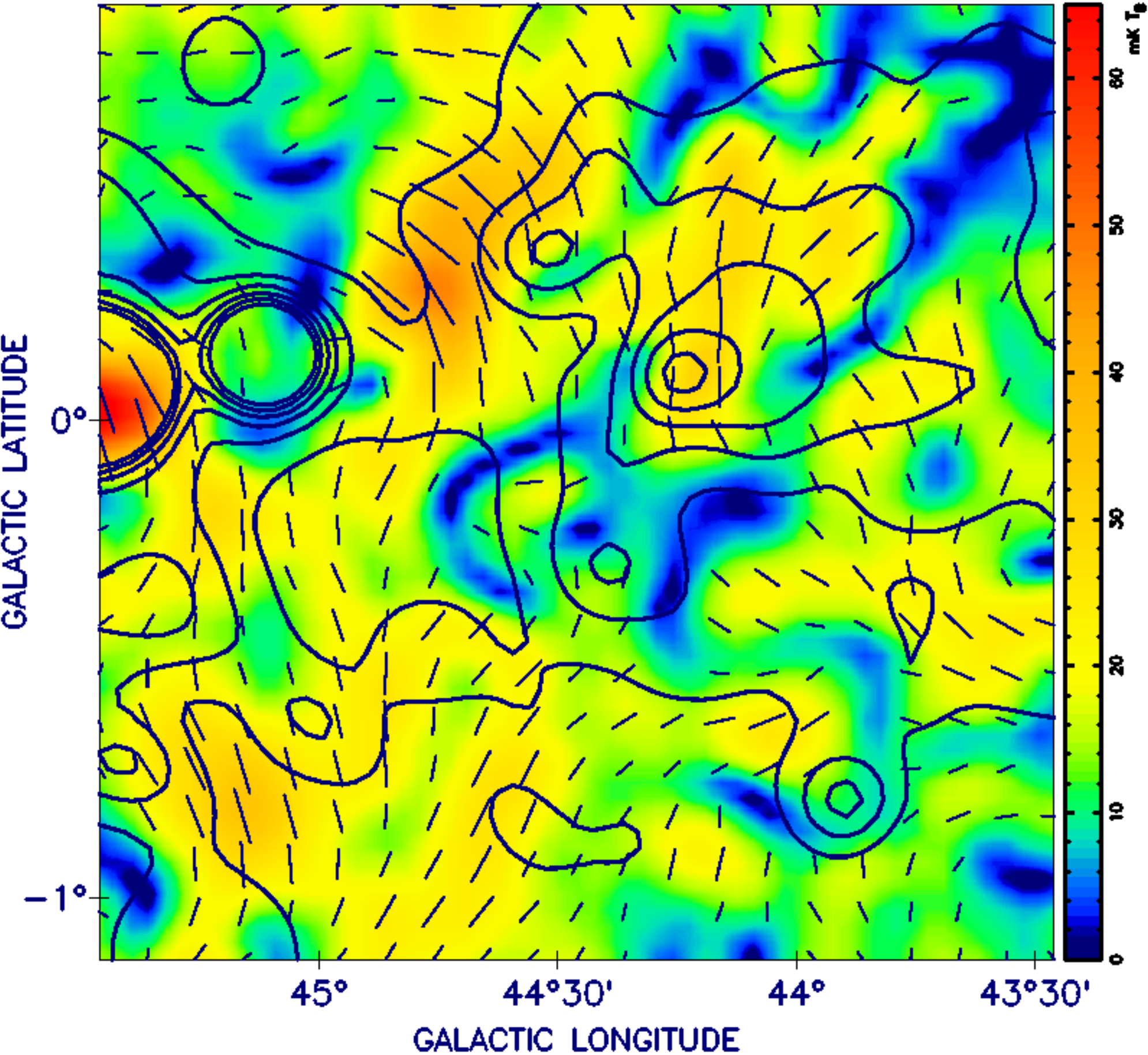}
\includegraphics[angle=0, width=0.5\textwidth]{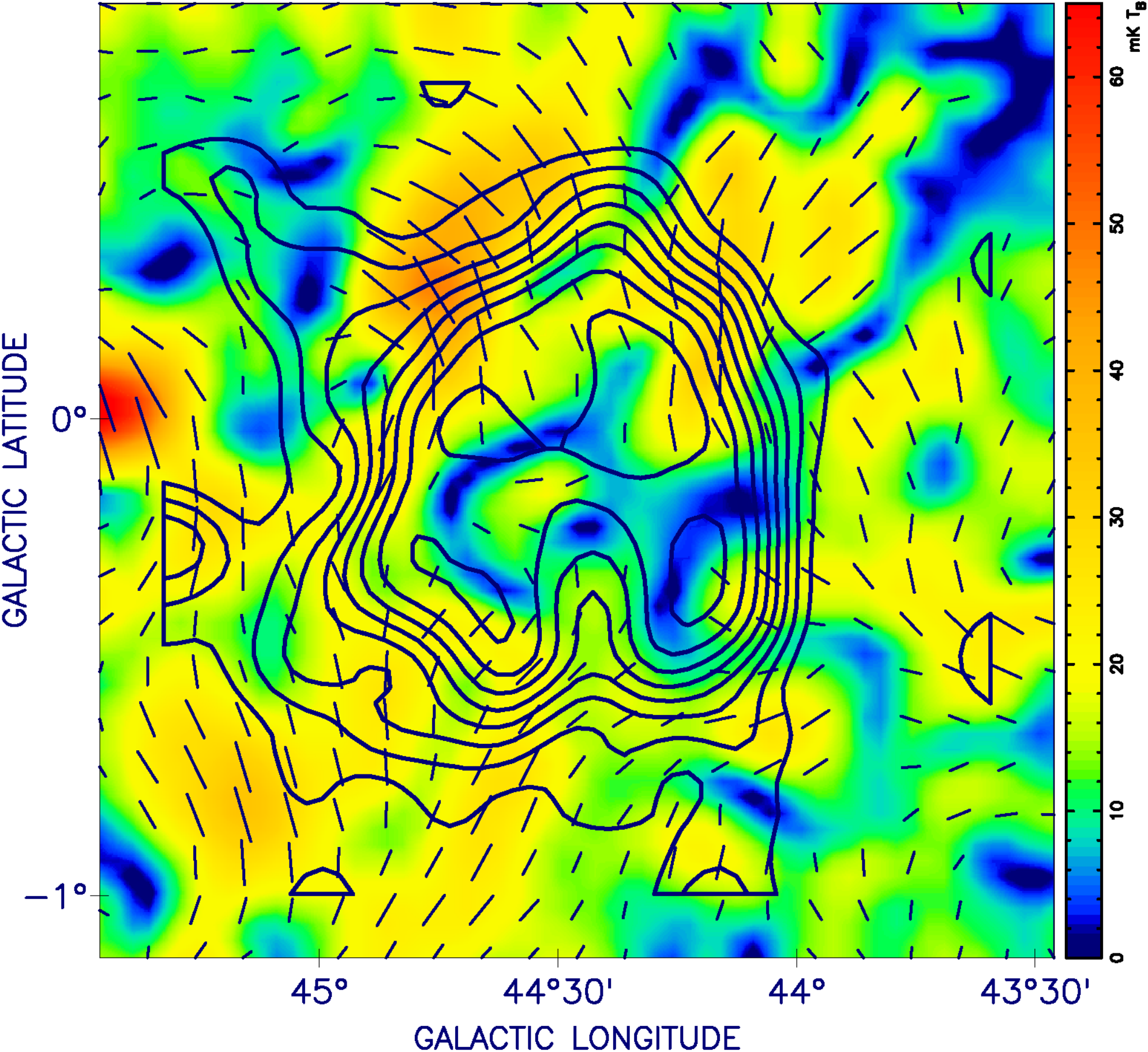}
\caption{{\it Left panel}: Effelsberg $\lambda$11\ cm polarised intensities with 
  overlaid vectors in B-field direction (for negligible Faraday rotation). Total-intensity 
  contours are 250~mK T$_{b}$ apart. 
  {\it Right panel}: Effelsberg $\lambda$11\ cm polarised intensities and vectors 
  as in the left panel overlaid with HESS J1912+101 TeV-emission contours as in 
  Fig.~\ref{6cm}.}
\label{11cm}
\end{figure*}

\subsubsection{$\lambda$6\ cm polarisation}

Figure~\ref{6cm} (left panel) shows radio emission details of the vicintity of 
HESS J1912+101 from the Urumqi $\lambda$6\ cm survey without large-scale emission
from WMAP. The direction of the polarisation 
vectors are in B-field direction (E-vectors $+90\degr$), which is correct for the case that 
Faraday rotation is negligible (see Sect. 3.3). The strong extended HII-region G45.1+0.1 
with a total integrated $\lambda$6\ cm flux density of about 6.9~Jy \citep{Reich14} shows 
some spurious instrumental polarisation at the periphery of the polarised shell. The 
distance of G45.1+0.1 was reported to be about 18~kpc \citep{Anderson09} and therefore
will not cause much depolarisation, because the polarised background beyond 18~kpc
is very faint. However, its instrumental polarisation confuses 
with the emission from the polarised arc eventually causing slight distortions.

In Figure~\ref{6cm} (right panel), we add $\lambda$6\ cm polarisation bars in B-field 
direction to polarised intensities and TeV-emission contours from HESS J1912+101. The 
HESS data were slightly convolved from $4\farcm2$ \citep{Puehlhofer15, HESS2018}  
to the angular resolution of the Urumqi survey of 9$\farcm$5. 
The polarised arc follows the outer TeV contours of HESS J1912+101
and overlaps with HESS J1912+101 in its southern area. The polarised signal reaches a 
maximum of nearly 30~mK (including diffuse Galactic emission), 
which clearly exceeds the typical observed polarised emission from this area of the
Galactic plane (see Fig.~\ref{6cmlarge}).

The $U$ and $Q$ maps need a correction to the local zero-level to integrate the emission
from the polarised arc. We have done this by subtracting $U$ and $Q$ emission
gradients across the field, then calculated the polarised emission and integrated it. 
The maximum polarised emission reduces to about 15~mK above the local zero-level. 
Because the zero-level and the emission boundaries are not exactly defined, the errors 
of the integrated polarised emission are significant. We measure  $0.5\pm0.2$~Jy 
for the polarised arc at $\lambda$6\ cm.  

\subsubsection{$\lambda$11\ cm polarisation}

In Fig.~\ref{11cm} (left panel), we show $\lambda$11\ cm polarisation intensities of 
the HESS J1912+101 area convolved to the $9\farcm5$ resolution of the 
Urumqi $\lambda$6\ cm map. This convolution 
increases the signal-to-noise ratio of the $\lambda$11\ cm data. The polarised emission 
at $\lambda$11\ cm disappears in the area of maximum polarisation at $\lambda$6\ cm 
and in the south, where the $\lambda$6\ cm polarised emission overlaps with 
HESS J1912+101 (Fig.~\ref{11cm}, right panel). This indicates higher depolarisation 
at $\lambda$11\ cm compared to $\lambda$6\ cm, which is expected.
The maximum polarised signal at $\lambda$11\ cm
at $4\farcm3$ resolution is about 60~mK and reduces to about 40~mK T$_{b}$ when convolved
to $9\farcm5$. 

\subsubsection{$\lambda$21\ cm polarisation}

\begin{figure*}
\includegraphics[angle=0, width=0.5\textwidth]{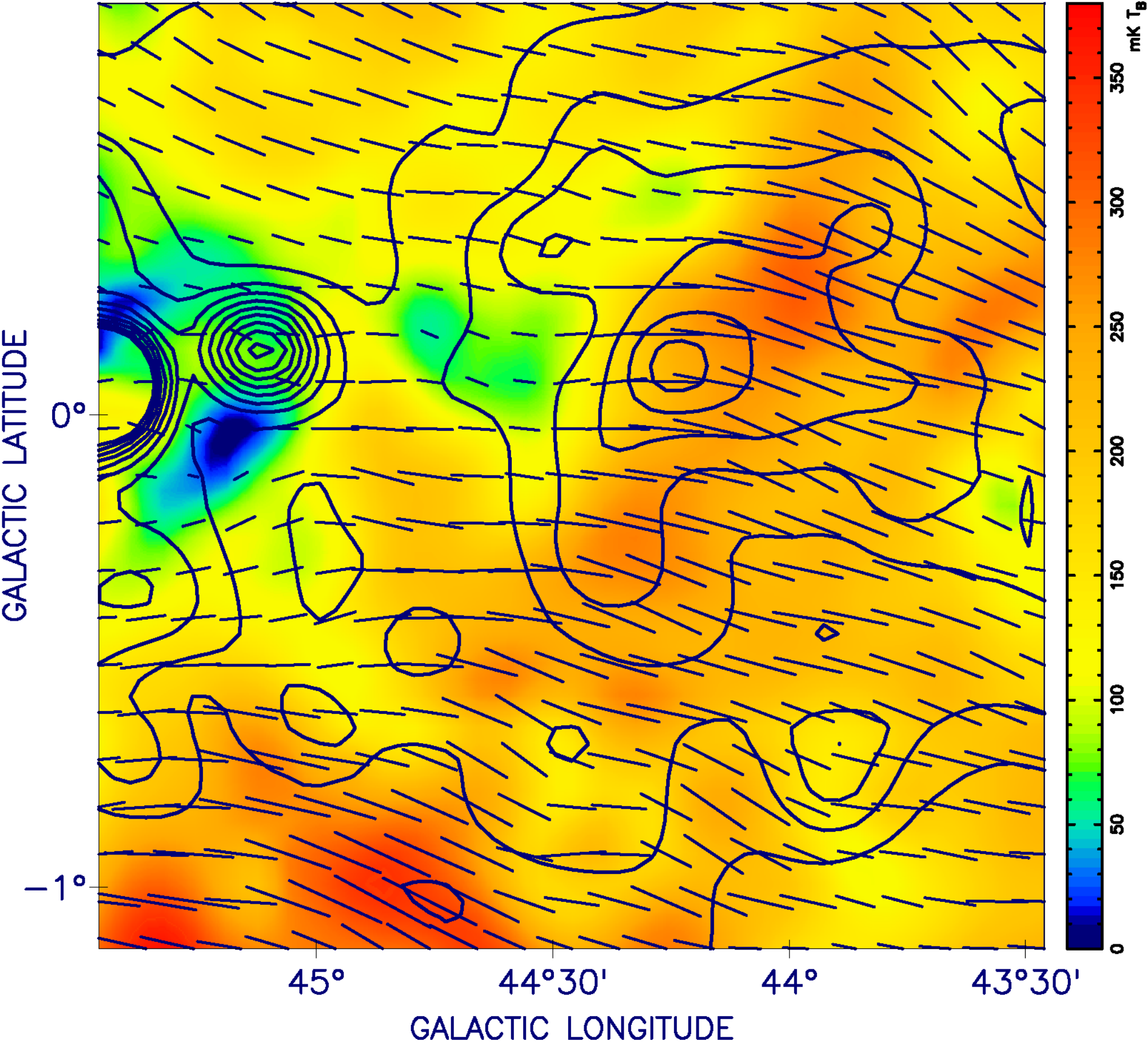}
\includegraphics[angle=0, width=0.5\textwidth]{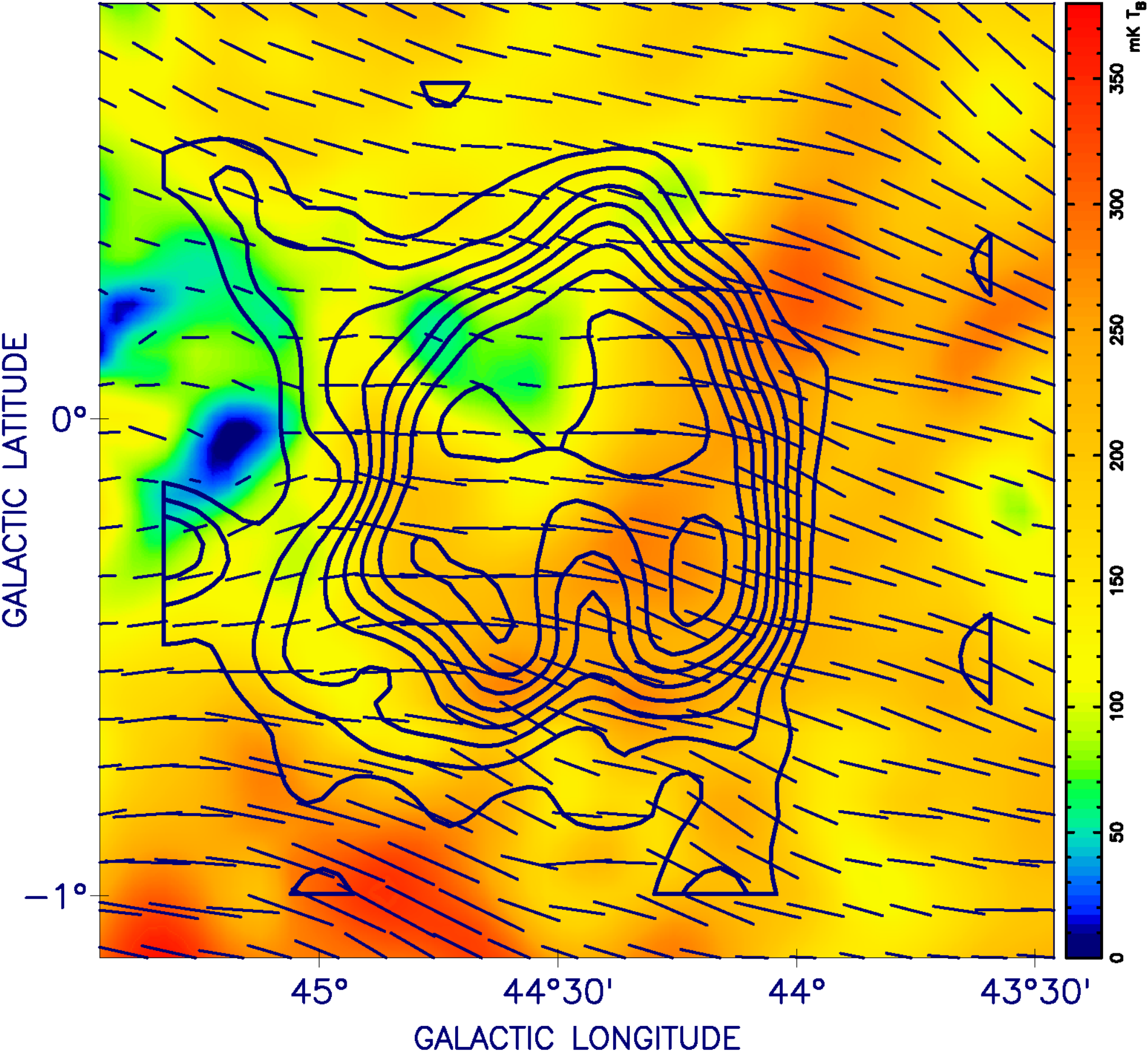}
\caption{{\it Left panel}: EMLS $\lambda$21\ cm polarised intensities with overlaid vectors 
  in B-field direction (for negligible Faraday rotation). Overlaid total intensity 
  contours are 1~K T$_{b}$ apart.
 {\it Right panel}: Effelsberg $\lambda$21\ cm polarised intensities and vectors 
  as in the left panel overlaid with HESS J1912+101 TeV-emission contours as in 
  Fig.~\ref{6cm}.}
\label{21cm}
\end{figure*}

The $\lambda$21\ cm emission distribution in the field of HESS J1912+101 
is shown in Fig.~\ref{21cm} (left panel). No correspondence between the 
polarised radio and the TeV emission is indicated (right panel). Distant $\lambda$21\ cm 
polarisation will be even 
more depolarised compared to $\lambda$11\ cm and we conclude that
the visible polarised emission originates entirely in the foreground of HESS J1912+101. 
We do not know the maximum distance up to which polarised emission at $\lambda$21\ cm 
is seen. A broad filament-like polarised intensity minimum at $\lambda$6\ cm running 
from about $l,b = 44\degr, 0\fdg2$ towards north-east seems to have an emission counterpart 
at $\lambda$21\ cm, indicating a complex radiation tranfer along the line-of-sight. 
This feature seems not to be related to HESS J1912+101.

\subsection{Total-intensity emission}

Both, the Urumqi $\lambda$6\ cm and the Effelsberg $\lambda$11\ cm total-intensity maps
reveal intense diffuse emission along the Galactic plane and also show
numerous strong slightly extended sources distributed around Galactic 
latitude $0\degr$. These maps do not show any extended total intensities 
obviously related to HESS J1912+101. As already mentioned, there is    
no signature of HESS J1912+101 in the high angular resolution $\lambda$21\ cm VLA 
surveys as well. 

\begin{figure*}[t]
\centering
\includegraphics[angle=0, width=0.2892\textwidth]{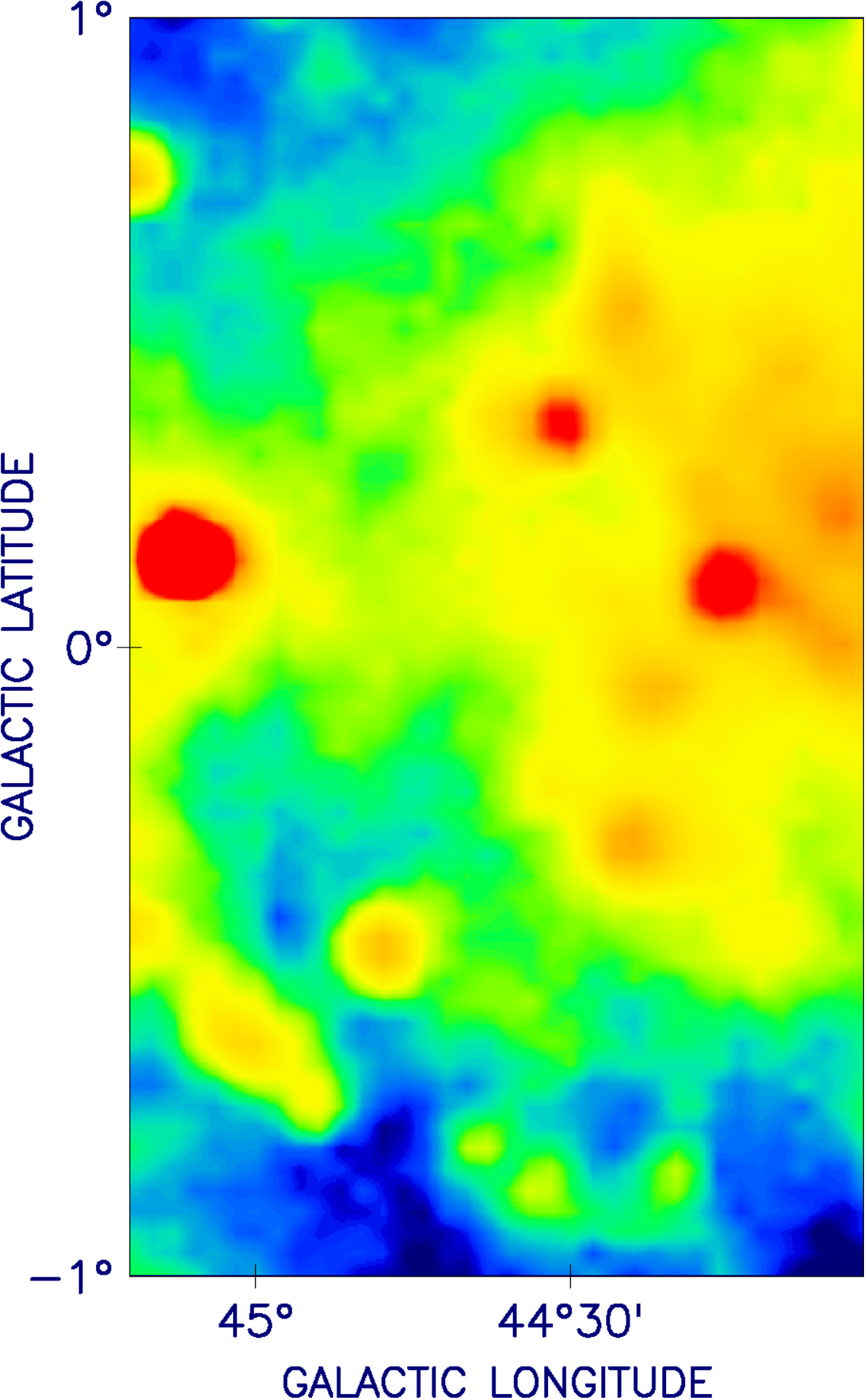}
\includegraphics[angle=0, width=0.2892\textwidth]{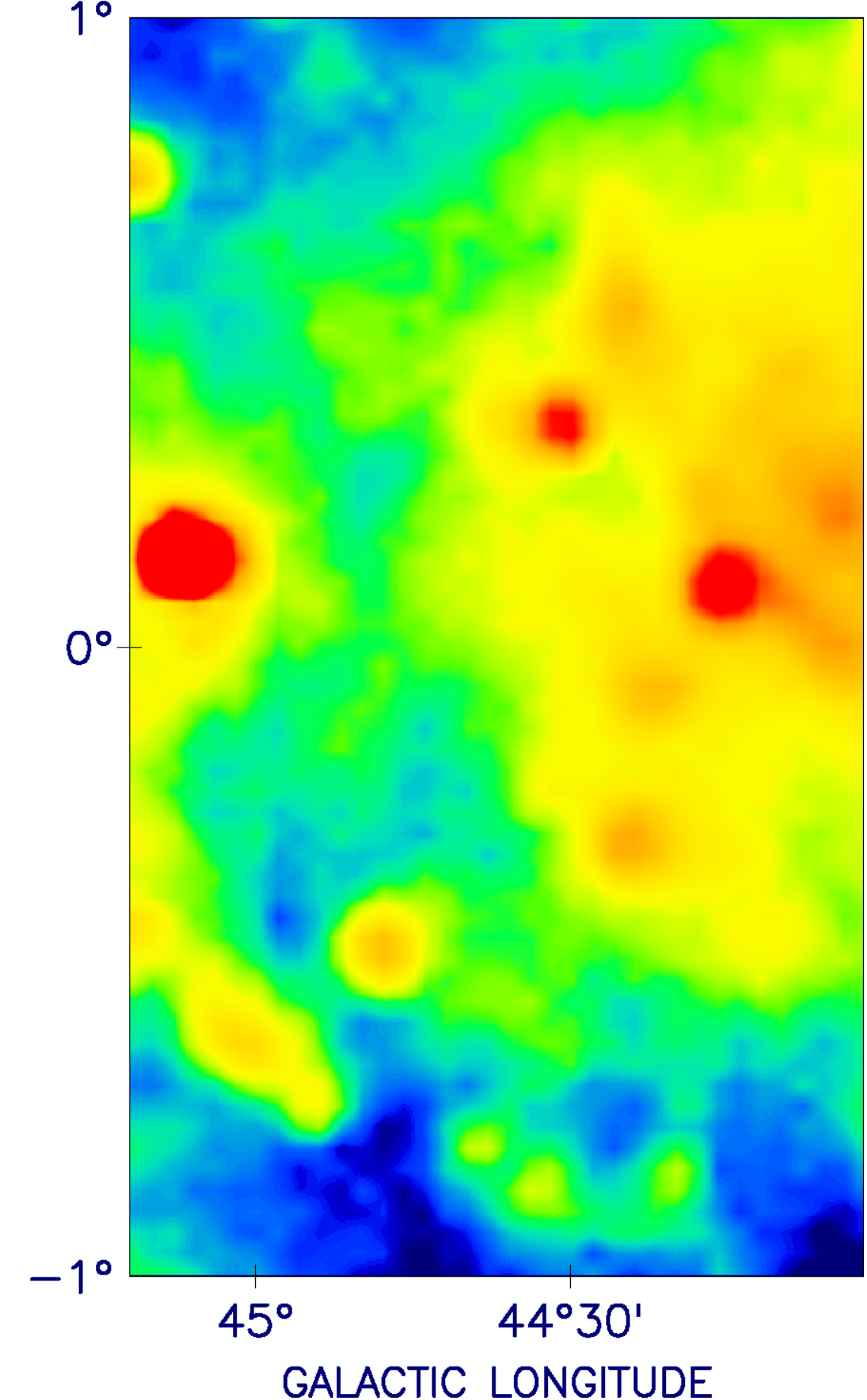}
\includegraphics[angle=0, width=0.3408\textwidth]{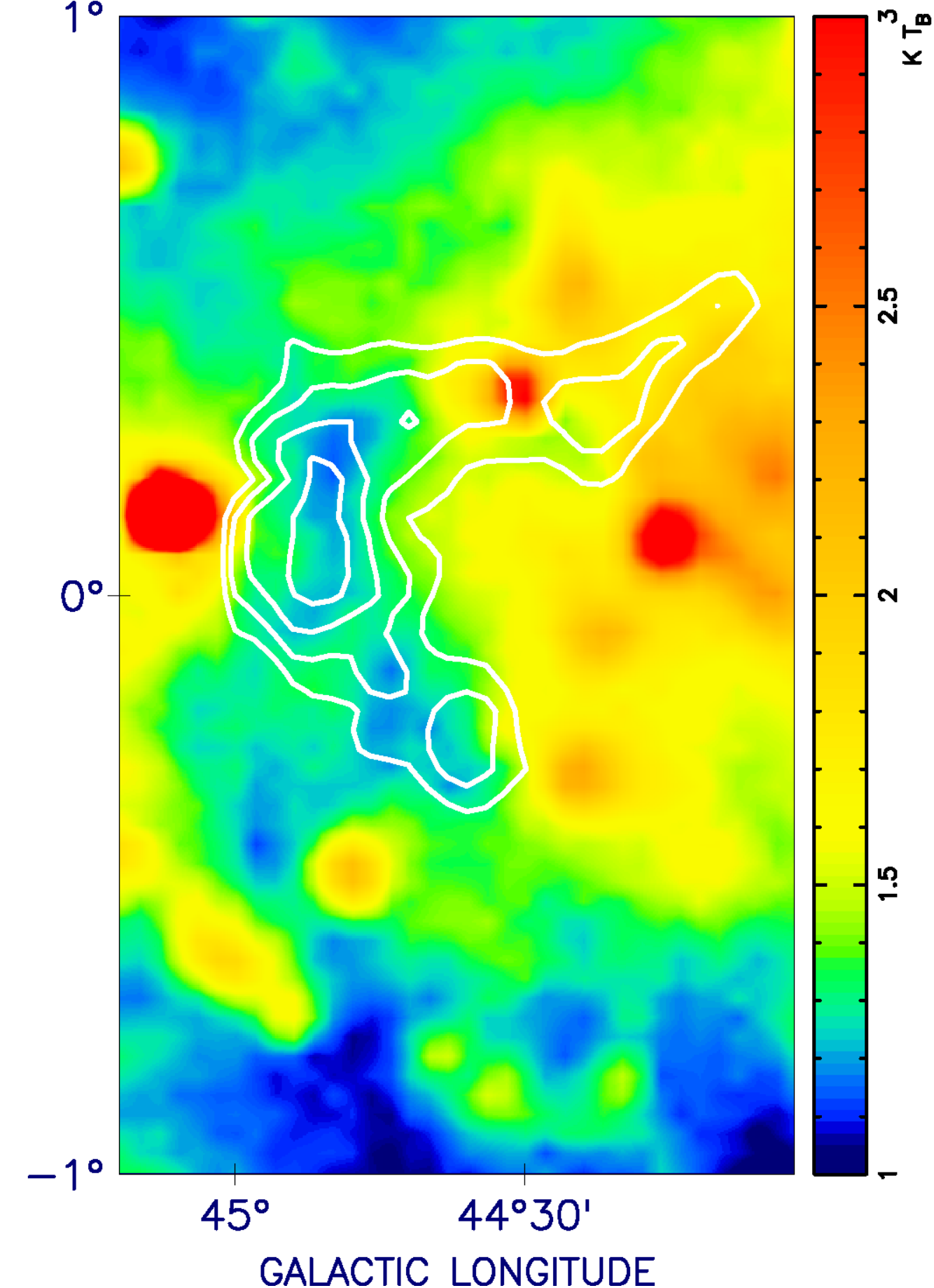}\\
\newpage
\caption{{\it From left to right}: Original $\lambda$11\ cm total-intensity map of 
 HESS J1912+101. {\it Middle panel} Model arc at $\lambda$11\ cm 
 for $40\%$ polarisation subtracted from the original. {\it Right panel} Same as in the
 middle panel, but for $20\%$ polarisation. The model arc was calculated from the 
 $\lambda$6\ cm polarised emission as indicated by contours.}
\label{11cm-PIshell}
\end{figure*}

As shown, the polarised $\lambda$6\ cm arc is from an emitting synchrotron 
source. We made attempts to identify the corresponding
total-intensity component, but were not successful. We may use indirect arguments
to constrain its properties by subtracting the scaled $\lambda$6\ cm polarised arc 
emission from total intensity maps to see its impact on the observed emission
by comparing with the emission after subtracting the total intensity model. We used the 
$\lambda$11\ cm  
total intensity map including large-scale background emission from the 
Stockert $\lambda$11\ cm 
survey \citep{Reif87} at its original resolution of 4$\farcm$3 in Fig.~\ref{11cm-PIshell}.
We subtracted the baselevel corrected $\lambda$6\ cm polarised arc scaled by a factor of 
10, which we calculate for a spectral index $\beta$ = -2.5 and $40\%$ polarisation,
and also a factor of 20 for the case of ${20\%}$ polarisation, from the 
total-intensity $\lambda$11\ cm map to compare the differences. 
The resulting maps are shown in Fig.~\ref{11cm-PIshell}. There is no indication of a 
distinct shell superimposed on large-scale diffuse emission, rather then a smooth 
depression of the diffuse total intensities. The thick arc merges with the more extended diffuse Galactic
emission. However, for polarisation fractions of 20\% and below, the model arc becomes
so strong in total-intensity that after subtraction minima in the survey map relative 
to the source-free areas at higher or lower latitudes results. This
is unlikely a real scenario. This way, we get a constraint of the arcs total-intensity 
contribution to the emission seen from this area.   

From the measured $\lambda$6\ cm polarised intensity of $0.5\pm0.2$~Jy, we estimate 
a total flux density for an average of $40(20)\%$ polarisation 
of $1.25\pm0.5(2.5\pm1.0)$~Jy at $\lambda$6\ cm for the area of the arc. The total
flux density from HESS J1912+101, however, may be higher as discussed in Sect.~4.

\subsection{Rotation measure}

We have calculated RMs from the $\lambda$11\ cm and $\lambda$6\ cm 
polarisation angles at $9\farcm5$ angular resolution. We are interested in the RM of 
the polarised arc and therefore used the $\lambda$11\ cm and $\lambda$6\ cm 
Stokes $U$ and $Q$ maps 
corrected for a local zero-level (see Sect.~3.1.2) 
before calculating polarisation angles. The results are influenced by depolarisation
at $\lambda$11\ cm in general, which is also reflected by unusual flat spectra 
for polarised synchrotron emission. For the $\lambda$11\ cm polarisation maximum, 
see Fig.~\ref{11cm}, we find a minimum RM of about +150 rad/m$^{2}$. 
The RM ambiguity of $\pm$371~rad/m$^{2}$ can not be neglected if data at 
only two wavelengths are available. 
Therefore, we have checked the ATNF Pulsar Catalogue \citep{man05}
\footnote{http://www.atnf.csiro.au/people/pulsar/psrcat/} for pulsars (PSRs) with measured 
RMs to compare with the RM values of the polarised arc. 
Unfortunately, there are only five PSR RMs listed within the $6\degr \times 6\degr$ field 
of Fig.~\ref{6cmlarge} showing a large scatter between +97 rad/m$^{2}$
up to +978 rad/m$^{2}$, which is not helpful for our case. However, all RMs are 
positiv. \citet{Xu14} list ten presumably extragalactic sources in the same field 
with RMs from the NVSS catalogue \citep{Taylor09} and the list of \citet{vanEck11}. 
These RMs show a similar large scatter as noted for the PSR RMs. Again, all sources 
have a positive sign, except for the faint source at $l,b = 46\fdg31,-0\fdg38$, 
where the NVSS RM of +529.2 rad/m$^{2}$ disagrees with the -117.0 rad/m$^{2}$ 
listed by \citet{vanEck11}. We conclude that the assumption of a positive RM for 
the diffuse polarised emission is justified. 
A RM of +150~rad/m$^{2}$ corresponds to a polarisation angle rotation of about 
34$\degr$ at $\lambda$6\ cm, which brings the polarisation vectors close to tangential 
to the HESS J1912+101 shell at the $\lambda$11\ cm polarisation maximum. For other parts 
of the arc, higher-frequency polarisation data are needed for a reliable estimate of RM 
and to find the intrinsic B-field direction of HESS J1912+101.

\section{Discussion}

\begin{figure} 
\centering
\includegraphics[angle=0, width=0.65\textwidth]{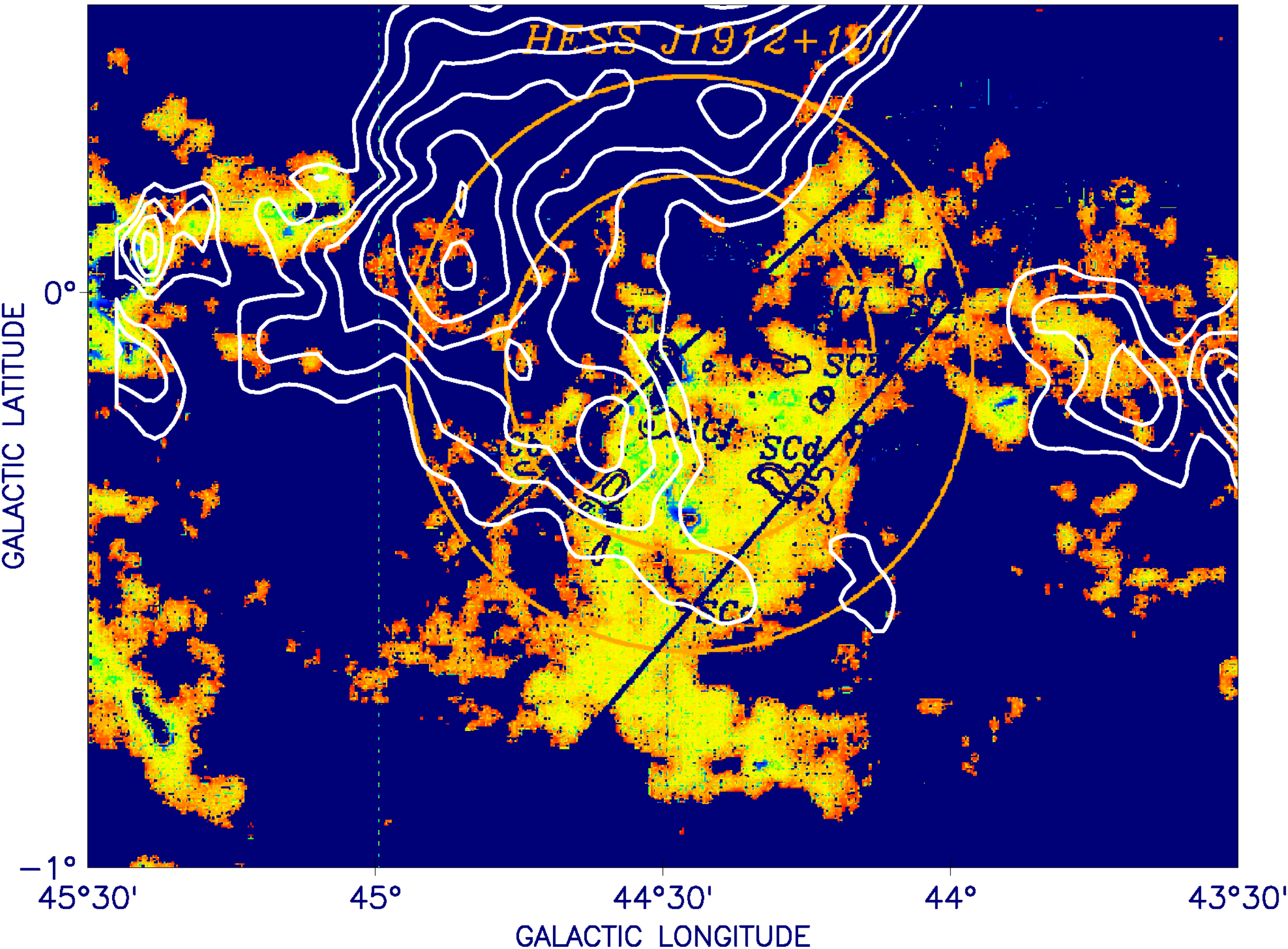}
\caption{Molecular gas distribution related to HESS J1912+101 taken from
 \citet{Su17} (their Fig.~4).
 The overlaid $\lambda$6\ cm PI contour lines are 3~mK T$_{b}$ apart
 and start at 12~mK T$_{b}$.}
\label{Su+PI}
\end{figure}

For HESS J1912+101, \citet{Su17} derived a distance of about 4.1~kpc based on HI 
self-absorption data, which is in rough agreement with the dispersion measure based 
distance of PSR J1913+1011 \citep{Morris02} of about 4.5~kpc, located near to the centre 
of HESS J1912+101.
The distance agreement suggests that PSR J19133+1011 and HESS J1912+101 are related. 
Also the age estimate by \citet{Su17} for HESS J1912+101 of about 0.7-2.0~$10^5$ years 
and that of PSR J193+1011 of about 1.7~$10^5$ years \citep{Morris02} agree within the 
errors and support a physical association. \citet{Su17} suggest a scenario, where the
stellar winds of the massive progenitor created a bubble in the molecular material.
This stellar wind also had an effect on the ambient magnetic field, which gets compressed. 
We note, that the polarised arc of HESS J1912+101 is located in an area with little
molecular material, when comparing with the CO-maps of \citet{Su17} (see Fig.~\ref{Su+PI}). 
Vice versa, 
areas with dense molecular clouds are not seen in polarisation. This fits into
the scenario that the polarised arc was formed by the stellar wind and subsequently
interacts with the SNR shock front. 
It is likely that the compression of the ambient magnetic field by the stellar wind
and/or the expanding SNR shell is distortet or suppressed in areas with clumpy molecular 
gas that large-scale polarisation is not observed. This will reduce the synchrotron 
emission from the SNR in this area. However, also enhanced depolarisation will result 
when thermal gas is associated with the molecular gas, which is currently unclear. 

The thermal HII-regions in the area of HESS J1912+101 for longitudes ${\le 44\fdg5}$ 
are all at larger distances \citep{Anderson12}, but we do not know the distance of the 
diffuse emission. In Sect.~3.2, we calculated the
total $\lambda$6\ cm flux density of the polarised arc based on an assumed percentage
polarisation, but this does not work for the other half of HESS J1912+101 dominated
by molecular clouds. For the same surface-brightness, the total flux density may be 
then be up to twice as large as estimated in Sect.~3.2, or about 2.5 to 5~Jy 
at $\lambda$6\ cm. Flux densities at 1~GHz range between 2.7 and 11.0~Jy depending
on the assumptions and for a typical SNR spectral index of $\alpha$ = -0.5 
($S \sim \nu^{\alpha}$). 
The estimated range of surface brightness values for HESS 1912+101 then is between
1.4~to~5.7$\times10^{-22}$~Wm$^{-2}$Hz$^{-1}$sr$^{-1}$. These values are about 
4 to 15 times larger than the surface brightness calculated above from the 
\citet{Su17} flux density estimate based on the work of \citet{Yamazaki06}, but are still 
very low. The spectral properties of old SNRs as described by \citet{Yamazaki06} are
basically in agreement with the radio data in view of the large uncertainties.

Even for the most conservative upper flux density estimate, the surface brightness of 
HESS 1912+101 of $\le~5.7\times10^{-22}$~Wm$^{-2}$Hz$^{-1}$sr$^{-1}$  belongs to
the 20\% known Galactic SNRs with the lowest surface-brightness (see Fig.~2 of
\citet{Case98}). SNRs with a low surface-brightness often show a high percentage
polarisation indicating a well-ordered magnetic field structure of their shells. 
Example SNRs were listed in Sect.~1.
These SNRs are evolved and have likely entered the cooling phase of SNR evolution. 
Their low surface-brightness makes it very unlikely that they can be detected in 
the Galactic plane, where diffuse synchrotron and thermal emission superimpose 
and where the density of Galactic
sources is high. In fact, most low surface-brightness SNRs are  located 
several degrees outside of the Galactic plane, where they just confuse with compact 
extragalactic sources but not with intense extended or structured emission. 
The polarised intensity signal from these SNRs is less confused,
but gets depolarised in the Galactic plane if they are too far away and the observing
frequency is not high enough that RM fluctuations cause beam 
depolarisation. The Urumqi $\lambda$6\ cm survey traces polarised Galactic
signals from large distances and clearly shows a partial polarised shell
surrounding HESS J1912+101.

\section{Summary}

We have 
shown that the polarised arc extracted from 
the Urumqi $\lambda$6\ cm and the Effelsberg $\lambda$11\ cm survey maps is 
very likely related to synchrotron emission from HESS J1912+101. This supports its 
identification as an old SNR. At $\lambda$11\ cm, the polarised emission is fainter 
because of depolarisation in view of high RMs, which are expected in the Galactic plane
for 4.1~kpc distance of HESS J1912+101 \citep{Su17}.   
We could not separate a total-intensity counterpart of HESS J1912+101 from 
confusing diffuse emission in the Galactic plane. SNRs of low-surface 
brightness have high-percentage polarisations, which allows to trace these objects by 
their polarised rather than by their total-intensity signal. HESS J1912+101 seems
to belong to this group of SNRs, where we estimate its percentage polarisation to
exceed 20\%. However, to fully settle the polarisation properties
of HESS J1912+101 and the intrinsic magnetic direction additional observations
at higher frequencies than 4.8~GHz with arcmin angular resolution are required.     

\begin{acknowledgements}

X.H.S. is supported by the National Natural Science Foundation of China under
grant no. 11763008. We like to thank Patricia Reich for careful reading of the 
manuscript and discussions and an anonymous reviewer for helpful comments.

\end{acknowledgements}

\bibliographystyle{raa}
\bibliography{bbfile}

\begin{thebibliography}{45}
\providecommand\natexlab[1]{#1}
\providecommand\JournalTitle[1]{#1}

\bibitem[{Aharonian} {et~al.}(2008)]{Aharonian08}
{Aharonian}, F., {Akhperjanian}, A.~G., {Barres de Almeida}, U., {et~al.} 2008,
  \aap, 484, 435

\bibitem[{Anderson} \& {Bania}(2009)]{Anderson09}
{Anderson}, L.~D., \& {Bania}, T.~M. 2009, \apj, 690, 706

\bibitem[{Anderson} {et~al.}(2012)]{Anderson12}
{Anderson}, L.~D., {Bania}, T.~M., {Balser}, D.~S., \& {Rood}, R.~T. 2012,
  \apj, 754, 62

\bibitem[{Bennett} {et~al.}(2013)]{Bennett13}
{Bennett}, C.~L., {Larson}, D., {Weiland}, J.~L., {et~al.} 2013, \apjs, 208, 20

\bibitem[{Case} \& {Bhattacharya}(1998)]{Case98}
{Case}, G.~L., \& {Bhattacharya}, D. 1998, \apj, 504, 761

\bibitem[{Duncan} {et~al.}(1999)]{Duncan99}
{Duncan}, A.~R., {Reich}, P., {Reich}, W., \& {F{\"u}rst}, E. 1999, \aap, 350,
  447

\bibitem[{Foster} {et~al.}(2013)]{Foster13}
{Foster}, T.~J., {Cooper}, B., {Reich}, W., {Kothes}, R., \& {West}, J. 2013,
  \aap, 549, A107

\bibitem[{Gao} {et~al.}(2011)]{Gao11y}
{Gao}, X.~Y., {Sun}, X.~H., {Han}, J.~L., {et~al.} 2011, \aap, 532, A144

\bibitem[{Gao} {et~al.}(2010)]{Gao10}
{Gao}, X.~Y., {Reich}, W., {Han}, J.~L., {et~al.} 2010, \aap, 515, A64

\bibitem[{Gorham}(1990)]{Gorham90}
{Gorham}, P.~W. 1990, \apj, 364, 187

\bibitem[{Gottschall} {et~al.}(2017)]{Gottschall17}
{Gottschall}, D., {Capasso}, M., {Deil}, C., {et~al.} 2017, in American
  Institute of Physics Conference Series, Vol. 1792, 6th International
  Symposium on High Energy Gamma-Ray Astronomy, 040030

\bibitem[{Green}(2017)]{Green17}
{Green}, D.~A. 2017, VizieR Online Data Catalog, 7278

\bibitem[{H.~E.~S.~S.~Collaboration} {et~al.}(2018)]{HESS2018}
{H.~E.~S.~S.~Collaboration}, {Abdalla}, H., {Abramowski}, A., {et~al.} 2018,
  \aap, 612, A8

\bibitem[{Han} {et~al.}(2015)]{Han15}
{Han}, J.~L., {Reich}, W., {Sun}, X.~H., {et~al.} 2015, Highlights of
  Astronomy, 16, 394

\bibitem[{Junkes} {et~al.}(1987)]{Junkes87}
{Junkes}, N., {F{\"u}rst}, E., \& {Reich}, W. 1987, \aaps, 69, 451

\bibitem[{Kassim}(1988{\natexlab{a}})]{Kassim882}
{Kassim}, N.~E. 1988{\natexlab{a}}, \apjl, 328, L55

\bibitem[{Kassim}(1988{\natexlab{b}})]{Kassim881}
{Kassim}, N.~E. 1988{\natexlab{b}}, \apjs, 68, 715

\bibitem[{Kothes} {et~al.}(2017)]{Kothes17}
{Kothes}, R., {Reich}, P., {Foster}, T.~J., \& {Reich}, W. 2017, \aap, 597,
  A116

\bibitem[{Li} {et~al.}(1991)]{Li91}
{Li}, Z., {Wheeler}, J.~C., {Bash}, F.~N., \& {Jefferys}, W.~H. 1991, \apj,
  378, 93

\bibitem[{Manchester} {et~al.}(2005)]{man05}
{Manchester}, R.~N., {Hobbs}, G.~B., {Teoh}, A., \& {Hobbs}, M. 2005, \aj, 129,
  1993

\bibitem[{Morris} {et~al.}(2002)]{Morris02}
{Morris}, D.~J., {Hobbs}, G., {Lyne}, A.~G., {et~al.} 2002, \mnras, 335, 275

\bibitem[{Puehlhofer} {et~al.}(2015)]{Puehlhofer15}
{Puehlhofer}, G., {Brun}, F., {Capasso}, M., {et~al.} 2015, in International
  Cosmic Ray Conference, Vol.~34, 34th International Cosmic Ray Conference
  (ICRC2015), 886

\bibitem[{Reich} \& {Reich}(1986)]{Reich86}
{Reich}, P., \& {Reich}, W. 1986, \aaps, 63, 205

\bibitem[{Reich}(1982)]{Reich82}
{Reich}, W. 1982, \aaps, 48, 219

\bibitem[{Reich} {et~al.}(1979)]{Reich79}
{Reich}, W., {Berkhuijsen}, E.~M., \& {Sofue}, Y. 1979, \aap, 72, 270

\bibitem[{Reich} {et~al.}(1992)]{Reich92}
{Reich}, W., {F{\"u}rst}, E., \& {Arnal}, E.~M. 1992, \aap, 256, 214

\bibitem[{Reich} {et~al.}(1984)]{Reich84}
{Reich}, W., {F{\"u}rst}, E., {Haslam}, C.~G.~T., {Steffen}, P., \& {Reif}, K.
  1984, \aaps, 58, 197

\bibitem[{Reich} {et~al.}(1990)]{Reich9011}
{Reich}, W., {F{\"u}rst}, E., {Reich}, P., \& {Reif}, K. 1990, \aaps, 85, 633

\bibitem[{Reich} {et~al.}(2004)]{Reich04}
{Reich}, W., {F{\"u}rst}, E., {Reich}, P., {et~al.} 2004, in The Magnetized
  Interstellar Medium, ed. B.~{Uyan{\i}ker}, W.~{Reich}, \& R.~{Wielebinski}
  (Copernicus GmbH, Katlenburg-Lindau), 45

\bibitem[{Reich} {et~al.}(2014)]{Reich14}
{Reich}, W., {Sun}, X.~H., {Reich}, P., {et~al.} 2014, \aap, 561, A55

\bibitem[{Reif} {et~al.}(1987)]{Reif87}
{Reif}, K., {Reich}, W., {Steffen}, P., {M{\"u}ller}, P., \& {Weiland}, H.
  1987, Mitteilungen der Astronomischen Gesellschaft Hamburg, 70, 419

\bibitem[{Su} {et~al.}(2017)]{Su17}
{Su}, Y., {Zhou}, X., {Yang}, J., {et~al.} 2017, \apj, 845, 48

\bibitem[{Sun} {et~al.}(2007)]{Sun07}
{Sun}, X.~H., {Han}, J.~L., {Reich}, W., {et~al.} 2007, \aap, 463, 993

\bibitem[{Sun} {et~al.}(2011)]{Sun11a}
{Sun}, X.~H., {Reich}, W., {Han}, J.~L., {et~al.} 2011, \aap, 527, A74

\bibitem[{Tammann} {et~al.}(1994)]{Tammann94}
{Tammann}, G.~A., {Loeffler}, W., \& {Schroeder}, A. 1994, \apjs, 92, 487

\bibitem[{Taylor} {et~al.}(2009)]{Taylor09}
{Taylor}, A.~R., {Stil}, J.~M., \& {Sunstrum}, C. 2009, \apj, 702, 1230

\bibitem[{Uyan{\i}ker} {et~al.}(1998)]{Uyaniker98}
{Uyan{\i}ker}, B., {F{\"u}rst}, E., {Reich}, W., {Reich}, P., \& {Wielebinski},
  R. 1998, \aaps, 132, 401

\bibitem[{Uyan{\i}ker} {et~al.}(1999)]{Uyaniker99}
{Uyan{\i}ker}, B., {F{\"u}rst}, E., {Reich}, W., {Reich}, P., \& {Wielebinski},
  R. 1999, \aaps, 138, 31

\bibitem[{Van Eck} {et~al.}(2011)]{vanEck11}
{Van Eck}, C.~L., {Brown}, J.~C., {Stil}, J.~M., {et~al.} 2011, \apj, 728, 97

\bibitem[{Wolleben} {et~al.}(2006)]{Wolleben06}
{Wolleben}, M., {Landecker}, T.~L., {Reich}, W., \& {Wielebinski}, R. 2006,
  \aap, 448, 411

\bibitem[{Xiao} {et~al.}(2011)]{Xiao11}
{Xiao}, L., {Han}, J.~L., {Reich}, W., {et~al.} 2011, \aap, 529, A15

\bibitem[{Xiao} {et~al.}(2009)]{Xiao09}
{Xiao}, L., {Reich}, W., {F{\"u}rst}, E., \& {Han}, J.~L. 2009, \aap, 503, 827

\bibitem[{Xu} \& {Han}(2014)]{Xu14}
{Xu}, J., \& {Han}, J.~L. 2014, Research in Astronomy and Astrophysics, 14, 942

\bibitem[{Xu} {et~al.}(2007)]{Xu07}
{Xu}, J.~W., {Han}, J.~L., {Sun}, X.~H., {et~al.} 2007, \aap, 470, 969

\bibitem[{Yamazaki} {et~al.}(2006)]{Yamazaki06}
{Yamazaki}, R., {Kohri}, K., {Bamba}, A., {et~al.} 2006, \mnras, 371, 1975

\end{thebibliography}

\end{document}